\documentclass[onecolumn,journal]{IEEEtran}
\usepackage{cite}

	\ifCLASSINFOpdf

   \usepackage[pdftex]{graphicx}
   \usepackage{epstopdf}
\else
   \usepackage[dvips]{graphicx}
\fi
\usepackage{amsmath}
\usepackage{enumitem}
\usepackage{color}
\usepackage{balance}
\usepackage{flushend}

\begin{document}
%
\title{Cooperation in 5G HetNets: Advanced Spectrum Access and D2D Assisted Communications}
%
%
%
\author{\IEEEauthorblockN{Georgios I. Tsiropoulos,~\IEEEmembership{Member,~IEEE},
Animesh Yadav,~\IEEEmembership{Member,~IEEE},
Ming Zeng,~\IEEEmembership{Student Member,~IEEE}, and
Octavia A. Dobre,~\IEEEmembership{Senior Member,~IEEE}}

\thanks{Georgios I. Tsiropoulos is with the School of Electrical \& Computer Engineering, National Technical University of Athens, Greece, GR15780 (e-mail:gitsirop@mail.ntua.gr).}

\thanks{Animesh Yadav, Ming Zeng and Octavia A. Dobre are with the Faculty of Engineering and Applied Science, Memorial University, St. John's, NL, A1B 3X5, Canada. (e-mail: \{animeshy, mzeng, odobre\}@mun.ca).}

}

\maketitle


\begin{abstract}
The evolution of conventional wireless communication networks to the fifth generation (5G) is driven by an explosive increase in the number of wireless mobile devices and services, as well as their demand for all-time and everywhere connectivity, high data rates, low latency, high energy-efficiency and improved quality of service. To address these challenges, 5G relies on key technologies, such as full duplex (FD), device-to-device (D2D) communications, and network densification. In this article, a heterogeneous networking architecture is envisioned, where cells of different sizes and radio access technologies coexist. Specifically, collaboration for spectrum access is explored for both FD- and cognitive-based approaches, and cooperation among devices is discussed in the context of the state-of-the-art D2D assisted communication paradigm. The presented cooperative framework is expected to advance the understandings of the critical technical issues towards dynamic spectrum management for 5G heterogeneous networks.
\end{abstract}


\section{Introduction}
Recently, Groupe Speciale Mobile Association has predicted that 8.9 billion connections (excluding machine-to-machine) will serve 5.6 billion unique subscribers by 2020 \cite{GSMA-2016}. To cater ever increasing demands for higher data rates in the fifth generation (5G) era and given that the wireless link efficiency is approaching its fundamental limit, researchers are seeking new paradigms to revolutionize the traditional communication technologies and employ unconventional thinking for 5G. The emerging networks are expected to be a mixture of nodes with different transmission powers and coverage sizes, as well as multiple tiers and different radio access technologies. Using technologies such as full-duplex (FD), device-to-device (D2D) and carrier aggregation (CA), as well as exploiting shared frequency bands belonging to the TV white spaces and radar, and unlicensed bands both below and above 6 GHz is envisioned. The 3rd Generation Partnership Project (3GPP) has defined the specification of the licensed-assisted access \cite{3GPP-RP-151569-2015} in Release 13 (LTE-Advanced Pro), to address the issue of unlicensed spectrum utilization by wireless networks while fairly coexisting with Wi-Fi and other technologies in the 5 GHz unlicensed band. The use of unlicensed bands above 6 GHz is considered in terms of specification initiatives for 5G New Radio (NR), which will become the foundation for the next generation of mobile networks. The NR standardization procedure is split into two phases: phase 1 for sub-40 GHz and phase 2 for sub-100 GHz, which will be defined in Releases 15 and 16, respectively. Accordingly, cooperation and coordination in terms of spectrum access play a vital access road for 5G network evolution. 

Cooperative communications facilitates the collaboration among terminals in a wireless network where nodes assist the information transmission of each other. Originally, relays were employed to realize the cooperative communications with an idea not only to overcome the path loss, but also to achieve diversity gain. However, the 5G vision recognizes that cooperation in future mobile wireless networks will not be strictly limited in assisting the information transmitted. Users may collaborate in terms of accessing the common wireless medium, exchanging information regarding the network and channels status, forming small groups where increased cooperative operations are applied to each other, and employing new technologies in favor of coordinated communication. 

From this perspective, in this article we investigate the cooperation in 5G heterogeneous network (HetNet), aiming to improve the system spectral efficiency (SE) while addressing the challenges of future mobile systems in terms of quality-of-service (QoS) provision. Our contributions can be summarized as follows:
\begin{itemize}
\item We present dynamic spectrum access (DSA) techniques that involve collaboration and interaction among users. Furthermore, we discuss the cooperative aspect of FD-based DSA, as well as explore cognitive-inspired spectrum access and CA techniques which enable more flexibility and improve spectrum utilization. Additionally, we examine recent advances for cooperative access of licensed bands focusing on spectrum trading techniques. 
\item We investigate the cooperation among devices in terms of D2D communications. We provide the taxonomy of cooperative D2D communications for 5G HetNet based on the level of centralization, use case scenarios and spectrum band under consideration. 
\item Different scenarios are considered for link establishment for the end-to-end connection in cooperative 5G networks. The impact of the signal-to-interference plus noise ratio (SINR) threshold and the distance between the nodes on network performance is studied in terms of ergodic throughput. Evidently, the distance among users may promote a certain cooperative option over the others, which in turn affects the throughput. 
\end{itemize}

The rest of this article is organized as follows. In Section II, we summarize the emerging trends in cooperative 5G HetNet from the perspective of spectrum access and D2D communications. In Section III, we describe diverse approaches to cooperative DSA, emphasizing on FD, CA, and cognitive-inspired methods. In Section IV, we present the state-of-the-art on D2D communications, and a potential cooperative communication model employing D2D and FD techniques is proposed in Section V. Finally, the paper is concluded in Section VI.

\section{Emerging Trends in Cooperative 5G HetNets}
Our view of cooperation in 5G HetNet is a collaborative and coordinated process among network nodes which takes advantage of the FD and D2D technologies, as well as the relay-assisted transmission, as summarized in Fig. 1. Furthermore, it can use either licensed or unlicensed, or shared spectrum bands. The basis and description of the network functionalities for cooperative processes in future 5G HetNet is presented below. 
\subsection{Prospects for Advanced Cooperative Spectrum Access}
It is evident that the currently, globally employed forty license frequency bands exclusive for cellular use are not enough to cater future traffic demands. Hence, 5G networks should be allowed to operate over unlicensed or shared spectrum bands. Shared bands in the TV white spaces (50-700 MHz) and radar bands (L/S/C bands) are among candidates. The unlicensed spectrum around 2.4 GHz, 5 GHz, and millimetre wave (mmWave) bands (57-64 GHz) represent potential bands to support 5G. Further, an SE improving CA technology is already employed in the licensed band in 3GPP Release 10, and has been extended to aggregation between carriers in both licensed and unlicensed spectrum bands in Release 13 \cite{3GPP-RP-151569-2015}. The application of CA in both uplink and downlink, along with users cooperation in terms of suitable carrier selection, may promote SE.

FD communications is another SE enhancing technology which has been considered as a promising candidate for 5G networks \cite{Zhang-other-CM-15}. The FD technology allows nodes to transmit and receive on the same frequency band simultaneously. FD communications can be incorporated with cooperative communications or/and underlay DSA techniques to improve spectrum utilization in 5G networks \cite{Kim-Lee-Hong-ST-2015,Gaafar-Other-TWC-2017}.

Advanced spectrum access models tackle the issue of sharing spectrum through an economic market or social perspective based on technical, business and social criteria. Licensed spectrum can be traded or leased among network nodes or tiers in a 5G HetNet \cite{Yang-Other-TWC-2016} to minimize the spectrum holes in time and frequency domains, and thus, increasing the system capacity.

\subsection{Device-to-Device Assisted Communications}
D2D communications in cellular networks is defined as direct or multihop communications between two incumbent users in the licensed or unlicensed, or shared bands without any or limited core network involvement. The cooperation among nearby devices is of considerable potential for network operators to offload traffic from the core network. At the same time, it represents a new communication paradigm to support social networking through localization. 

D2D assisted communications is one of the highlights of the 3GPP Releases 12 and 13 \cite{3GPP-RP-151569-2015}. Devices will cooperate with each other operating as mobile relays, and exploiting their spatial diversity advantages. Moreover, D2D functionality and user cooperation are considered not only to improve SE, but also to address other potential use cases, e.g., localization, multicasting, video dissemination, and machine-to-machine communications. Evidently, this reduces the cost of communication by potentially improving the energy efficiency (EE), delay, fairness and facilitates resource sharing.

\section{Diverse Approaches to Cooperative Dynamic Spectrum Access}
In this section, we discuss the way that cooperation among network nodes will revolutionize the DSA techniques.
\subsection{Cooperative Full-Duplex Spectrum Access Techniques}
The FD technology not only nearly doubles SE, but also reduces end-to-end delays and facilitates the evolvement of conventional half-duplex (HD) technology into a denser heterogeneous networking environment with mobile relaying nodes, as shown in Fig. 2(a). Because of the loop self-interference (SI), this technology can be practically feasible for short-range communication scenarios such as small cell base station (BS) and relay nodes. These nodes cooperate in the sense that while they exchange data, they acquire the channel state information in real-time, as shown in Figs. 2(b), (c). Moreover, there is a practical implementation of cooperation in terms of spectrum since downlink and uplink are transmitted over the same channel at the same time. Considering cognitive radio (CR)-based communication networks, the FD technology is exceptionally useful since cognitive users can monitor the licensed users activities, improve spectrum hole identification accuracy and reduce the probability of false alarm.

Cooperative FD communication can be extensively applied in relay-assisted wireless networks. Specifically, dedicated relays or users with good channel condition act as relays to assist the users with bad channel conditions \cite{Zhong-Zhang-COMML-16}. As illustrated in Figs. 2(d), (e), in the HD based wireless systems, a two-phase cooperation protocol is applied to realize the relay-assisted communication in time and frequency domains. Particularly, in phases I and II, the relay node listens and forwards the traffic of the user, respectively. The use of FD technology reduces the two-phase cooperation into one.

Recently, a new cooperative paradigm employing FD communication has been introduced in CR networks \cite{Zheng-Krikidis-Ottersten-twc-13}. The cognitive users may act as relays to forward the licensed users signal, and in return, they can utilize the licensed spectrum to serve their own communication needs. The conventional HD-based cooperation scenario employing multiple-input multiple-output (MIMO) is realized in two phases, as shown in Figs. 2(f), (g): 1) the cognitive user receives the signal from the licensed user, and 2) the cognitive user forwards the primary traffic and transmits its own signal at the same time. In the envisioned FD-based cooperative 5G networks, both phases can be realized simultaneously (Fig. 2(h)).

\subsection{Cognitive-inspired Spectrum Access}
The licenced spectrum policies limit the flexibility of spectrum utilization, while this dynamically changes over different geographical areas and time \cite{Hong-other-CM-14}. By means of 5G HetNet, the CR technology may be employed in the following aspects:
\begin{enumerate}[label=(\alph*)]
\item Allowing a smooth migration from the legacy radio access networks (RANs) to the new 5G communication era; BSs can dynamically lease channels from the previous generation RANs through the CR technology. 
\item Allowing small cells to opportunistically use a channel for a certain time to serve an incoming user. 
\item Allowing D2D to opportunistically occupy the licenced spectrum based on its availability over time and space \cite{Diamantoulakis-other-TVT-16}. 
\end{enumerate}

To promote the CR technology application to 5G HetNet, spectrum sensing and mobility can play a vital role. However, the radio sensing process can be complex and expensive, and may not perform satisfactorily with single user information. Hence, cooperation can enhance the sensing performance via spatial diversity and reduce the detection time at each individual user \cite{Chen-Chen-Meng-ST-14}. Furthermore, clustering of cognitive users based on their spatial distribution can  reduce the energy consumption and overhead due to the sensing information exchange, as the users within the same group have similar sensing characteristics. On the other hand, neighbouring groups of nodes can cooperate in terms of timely exchanging information with a low overhead. Moreover, node grouping can facilitate the frequency reuse among those that are spatially remote and alleviate the data collisions. The leased bands may be used to accommodate the overload traffic when the network operates near congestion or to provide opportunistic services at low cost. 

CR can be combined with CA techniques to allow the usage of multiple spectrum segments and exploit every slice of the frequency band \cite{Diamantoulakis-other-TVT-16}. Cooperation among nearby BSs facilitates a suitable selection of component carriers to mitigate the inter-cell interference. Since a communication path in 5G HetNet may not include the BS, the component carrier selection can be realized at the user devices. To this end, coordination between users is vital to ensure reliable communication. 

\subsection{Cooperative Access of Licenced Band: The Economic and Social-based Perspective of Dynamic Spectrum Sharing}
To design an efficient and effective DSA, an economic marketing perspective is also considered, which is referred to as spectrum trading or leasing. Two examples of cooperative-based architectures that realize dynamic spectrum trading techniques are as follows: 
\begin{enumerate}[label=(\alph*)]
\item A multi-tier architecture is considered where the spectrum owner, primary service provider (PSP), secondary service provider (SSP) and end users are the trading entities. Such an architecture can be used in HetNet, where the spectrum owner, PSP and SSP are the wireless operator, macrocell and small cell, respectively \cite{Yang-Other-TWC-2016}. 
\item	The licensed users may directly decide to lease the spectrum access rights or parts of it to cognitive users. The profit of the licensed users may be either economic revenue or technical cooperation \cite{Liu-Others-COMML-13}. In the latter case, the cognitive users act as relays in exchange for a fraction-of-time on accessing the licensed spectrum. Likewise, a trade-off results for cognitive users consuming the power on the relay transmission and acquiring sensing information. 
\end{enumerate}
Apart from cooperative architectures, cooperative policies based on user credits can be applied. Users can earn credits when they operate in a collaborative manner. User credits can be spent to gain cooperation from other users. 

Emerging cooperative-based architectures can exploit the information of social networks to improve their performance and facilitate information dissemination \cite{Zhang-Others-CM-17}. To this end, users with low social distance are expected to interact with increased frequency and are likely to be physically close. This information can be used by the D2D technology to form direct links among users. Moreover, socially close users are considered likely candidates to serve as relays in relay-assisted communications.
\subsection{Inband or Outband, Above or Below 6 GHz?}
The user can choose either inband (licensed band) or outband (shared/unlicensed band) spectrum for communications. Inband spectrum, which is fully controlled by the BS or controller, can be accessed via centralized or distributed fashion. In the centralized case, devices cooperate with the controller for channel availability, the QoS is guaranteed and interference management is easy. Contrastingly, in a distributed case, devices need to sense the wireless environment to use the channel. Neighbouring devices may cooperate, facilitating sensing information exchange to increase spectrum sensing efficiency. The main challenges of distributed inband communications are the interference management and resource allocation, which usually require methods of high complexity. An inadequate interference management may waste cellular resources and deteriorate the conventional cellular communication.

In unlicensed bands, there are no exclusive rights for their use, and therefore, they may be employed by any independent device that is compliant with the usage rules, such as maximum power levels and bandwidth limitations. It does not interfere the cellular network users; however, interference within the band is uncontrollable. Besides, security and QoS are not guaranteed. On the other hand, in shared bands, non-incumbent users are allowed to use the spectrum in accordance with sharing rules to avoid or limit the interference to the incumbent users. The most popular use cases of inband and outband communications are shown in Fig. 4(a). The spectrum below 6 GHz is generally recommended for long-distance communications. For instance, the IEEE 802.22 standard (shared band in the TV frequency spectrum) is used to provide broadband wireless access in rural areas for ranges up to 100 km. The spectrum bands above 6 GHz are useful for mid- and short-range links. For instance, the IEEE 802.11ad standard recommends the use of the unlicensed 60 GHz band for short distance, especially for indoor use achieving multi-gigabit speeds. The standardization bodies and the operators push the technology forward to use the above 6 GHz band through the upcoming 5G NR standards. In 5G NR, the combination of the two bands will be able to meet the wide range of 5G requirements.

\section{D2D Communications}
The D2D communications is considered from three perspectives: control plane, type of communication, i.e., information relay or direct communication, and application of FD technology, as depicted in Fig 3. 

\subsection{Level of Centralization: Core Network-Assisted versus Autonomous}
Devices may be partially or fully assisted by the BSs or relay nodes of the core network to control the D2D connections, as depicted in Fig. 4(b). The core network-assistance improves the throughput, EE and SE performances at the cost of extra computational and controlling load for BSs and relay nodes. Evidently, such a scenario is not beneficial for D2D deployments that involve out-of-core network coverage.

In autonomous D2D communications, devices are fully responsible to set up, control and coordinate the communication. Cooperation among devices is essential to allow distributed users process and relay information in a coordinated fashion. Channels belonging to different bands, i.e., licensed or unlicensed, or shared may be employed for the multiple hops, with different access methods, e.g., spectrum leasing or cognitive access. If the distance between communicating devices is small, they may prefer to select channels from bands above 6 GHz. In higher frequency bands, signal power attenuates significantly, and thus, the interference is restricted within a small region, which can be managed by applying the simple frequency division multiplexing technique within the region. It is worth mentioning that autonomous D2D systems are critical in natural disasters, such as earthquakes or hurricanes, since they can set up an urgent communication network replacing the damaged core network.
 
\subsection{D2D Use Case Scenarios: Direct or Relay Communication}
The introduction of D2D functionality in 5G HetNet creates two main use case scenarios, as illustrated in Fig. 4(c). In the first scenario, nearby devices may set up local links among them, which can potentially improve the user experience in terms of latency and power consumption, and lead to increased SE and EE by dense spectrum reuse. Multiple D2D links can operate over the same channel within the same cell, which increases spectrum reuse per cell beyond one. A challenging case is when multihop D2D communication links are involved, i.e., more than two devices cooperate to establish an end-to-end communication path. 

In the second scenario, a device with better transmission characteristics to the BS can act as a relay to assist the communication of nearby devices. Cooperative relaying D2D communications can benefit from the spatial diversity of devices, exploit multiuser shadow-diversity and improve transmission range. D2D relay introduces new technical challenges, where devices should collaborate to discover the candidate relays, select the relay device and minimize the power consumption of the relay device. It is envisioned that devices located within a cell can be clustered into several groups. A certain relay node aggregates the traffic from the devices within each group and applies multiplexing techniques to forward the traffic to the BS. As the transmission characteristics vary over time, the relay node selected within the group can change accordingly.

\subsection{FD D2D-aided Cooperative NOMA}
The FD technology can be combined with non-orthogonal multiple access (NOMA) in terms of D2D communications to further improve the SE. In cooperative NOMA relay-assisted CR networks, cognitive users help the licensed ones by acting as relays, while at the same time they use the same spectrum band to transmit their messages exploiting the NOMA technology advantages \cite{L_Lv-Others-TVT-16}. In another scenario, the NOMA user with higher SINR can receive data from the BS and forward them towards the NOMA user with lower SINR employing the FD technology at the same time over the same channel \cite{Z_Zhang-Others-TVT-16}. The application of FD D2D-aided cooperative NOMA can significantly improve the downlink performance and the transmission reliability of the NOMA user with lower SINR. In a novel use case, a BS can communicate to its low SINR user via a relay device, which employs NOMA to forward the data to the BS user and its own destination user; SE can be further improved by using the FD relay.

\section{Performance Studies}
In this section, we present illustrative performance studies for cooperative 5G HetNet employing D2D and FD technologies. Four scenarios are set to reflect the different alternatives offered in 5G HetNet to establish a link between two nodes, as depicted in Fig. 5(a). The spectrum below 6~GHz is considered. Two performance metrics, i.e., ergodic throughput versus SINR and D2D distance, are respectively studied to illustrate the performance of the algorithm described in Fig.~5(b).

When the transmitter and receiver devices are close enough, then an outband direct link is established, which corresponds to Scenario A. If the direct link quality is not adequate to satisfy the QoS requirements, then more than two devices cooperate to establish a communication path, i.e., Scenario B. If a link is not feasible without the core network's involvement, then Scenario C is used, which corresponds to the conventional link. D2D communications can be employed for distant users or users with poor communication characteristics to assist their communication with the BS, as considered in Scenario D. For Scenarios A and B, autonomous D2D links on the outband spectrum are used, which not only avoids the interference but also improves the SE. On the other hand, inband is employed for Scenarios C and D since they involve the core network. The 5G HetNet admits an incoming call request by the source device based on the SINR of the link, the QoS requirements and the potential availability of a relay device, as shown in Fig. 5(b). 

We assume a circular cell of radius $r=1$ km with a macro BS at the center. The locations of the inband and outband D2D transmitters are modelled via a Poisson point process with densities $\lambda_{in}$ and $\lambda_{out}$, respectively. A list of parameters and their values are summarized in Fig.~6(a).
In Fig. 6(b), the ergodic throughput in b/s/Hz versus the SINR is shown for the four scenarios. Scenarios B and D, which involve device cooperation, use the amplify-and-forward (AF) protocol. Scenarios C and D, which involve BS cooperation, use the decode-and-forward (DF) protocol. In summary, autonomous D2D links, such as Scenarios A and B obtain higher ergodic throughput when compared to the core network-assisted Scenarios C and D. Moreover, the FD mode further improves the SE as achieved by Scenarios B and D. It is worth mentioning that Scenario D with HD mode obtains the lowest performance because of the three-hop link. However, the main advantages of Scenarios C and D are link reliability and QoS guarantees for the entire duration of the session.

In Fig. 6(c), we study the effect of varying the D2D distance on the ergodic throughput of the four considered scenarios. The D2D link required SINR is fixed to 10 dB in this case. The throughput performance of all scenarios, except for C, decreases with the increase in the distance between two communicating devices. In Scenario C, the distance between the device to BS is kept fixed to values equal to 100 and 250 m, respectively. Further, owing to the cooperation in Scenario B, its performance is best among the considered scenarios. Leveraging our preliminary studies, different use case scenarios for cooperative DSA can be further explored towards 5G networks.

\section{Conclusion and Research Directions}
This article provided a glimpse of cooperative communications in 5G HetNet. Cooperation aims to improve the network performance in terms of SE, latency, connectivity, EE and QoS. Cooperation among the wide array of spectrum available in the shared and unlicensed bands, along with the link distance-dependent access for enhancing the SE are beneficial. Furthermore, employing FD and D2D technologies facilitates the cooperation among the distributed devices and cognitive users, and lead to an efficient DSA. Cooperative communications can be combined with 5G NR to get the most out of every bit of diverse spectrum. The insights gained from the preliminary numerical investigations regarding the potential adoption of cooperative techniques demonstrate the significant improvement in terms of SE, which is essential to meet the unprecedented challenges of future mobile systems.  

This article also paves the way for future research directions, as follows: First, it is important to optimize the various cooperative alternatives available in various bands; Second, higher device density causes higher aggregate interference, and thus, the need for efficient mitigation techniques is inevitable, especially for shared and unlicensed bands; Finally, the number of collisions increases with load in the shared and unlicensed bands, and hence, a clever spectrum coordination scheme should be devised to reap the full potential of these bands for cellular usage.

\section*{Acknowledgment}
This work has been supported in part by the Natural Sciences and Engineering Research Council of Canada (NSERC), though its Discovery program.

\bibliographystyle{IEEEtran}

\begin{thebibliography}{10}
\providecommand{\url}[1]{#1}
\csname url@samestyle\endcsname
\providecommand{\newblock}{\relax}
\providecommand{\bibinfo}[2]{#2}
\providecommand{\BIBentrySTDinterwordspacing}{\spaceskip=0pt\relax}
\providecommand{\BIBentryALTinterwordstretchfactor}{4}
\providecommand{\BIBentryALTinterwordspacing}{\spaceskip=\fontdimen2\font plus
\BIBentryALTinterwordstretchfactor\fontdimen3\font minus
  \fontdimen4\font\relax}
\providecommand{\BIBforeignlanguage}[2]{{%
\expandafter\ifx\csname l@#1\endcsname\relax
\typeout{** WARNING: IEEEtran.bst: No hyphenation pattern has been}%
\typeout{** loaded for the language `#1'. Using the pattern for}%
\typeout{** the default language instead.}%
\else
\language=\csname l@#1\endcsname
\fi
#2}}
\providecommand{\BIBdecl}{\relax}
\BIBdecl

\bibitem{GSMA-2016}
{GSMA Intelligence}, ``The mobile economy,'' Tech. Rep., 2016.

\bibitem{3GPP-RP-151569-2015}
3GPP, ``Release 13 analytical view,'' RP-151569, Tech. Rep., Sep. 2015.

\bibitem{Zhang-other-CM-15}
Z.~Zhang \emph{et~al.}, ``Full-duplex techniques for 5{G} networks:
  {S}elf-interference cancellation, protocol design, and relay selection,''
  \emph{{IEEE} Commun. Mag.}, vol.~53, no.~5, pp. 128--137, May 2015.

\bibitem{Kim-Lee-Hong-ST-2015}
D.~Kim, H.~Lee, and D.~Hong, ``A survey of in-band full-duplex transmission:
  {F}rom the perspective of {PHY} and {MAC} layers,'' vol.~17, no.~4, pp.
  2017--2046, 4th Quart., 2015.

\bibitem{Gaafar-Other-TWC-2017}
M.~Gaafar \emph{et~al.}, ``Underlay spectrum sharing techniques with in-band
  full-duplex systems using improper {G}aussian signaling,'' \emph{{IEEE}
  Trans. Wireless Commun.}, vol.~16, no.~1, pp. 235--249, Jan. 2017.

\bibitem{Yang-Other-TWC-2016}
C.~Yang \emph{et~al.}, ``Advanced spectrum sharing in 5{G} cognitive
  heterogeneous networks,'' \emph{{IEEE} Trans. Wireless Commun.}, vol.~23,
  no.~2, pp. 94--101, May 2016.

\bibitem{Zhong-Zhang-COMML-16}
C.~Zhong and Z.~Zhang, ``Non-orthogonal multiple access with cooperative
  full-duplex relaying,'' \emph{{IEEE} Commun. Lett.}, vol.~20, no.~12, pp.
  2478--2481, Dec. 2016.

\bibitem{Zheng-Krikidis-Ottersten-twc-13}
G.~Zheng, I.~Krikidis, and B.~O. Ottersten, ``Full-duplex cooperative cognitive
  radio with transmit imperfections,'' \emph{{IEEE} Trans. Wireless Commun.},
  vol.~12, no.~5, pp. 2498--2511, May 2013.

\bibitem{Hong-other-CM-14}
X.~Hong \emph{et~al.}, ``Cognitive radio in 5{G}: {A} perspective on
  energy-spectral efficiency trade-off,'' \emph{{IEEE} Commun. Mag.}, vol.~52,
  no.~7, pp. 46--53, Jul. 2014.

\bibitem{Diamantoulakis-other-TVT-16}
P.~Diamantoulakis \emph{et~al.}, ``Carrier aggregation for cooperative
  cognitive radio networks,'' \emph{{IEEE} Trans. Veh. Technol.}, no.~99, pp.
  1--1, Dec. 2016.

\bibitem{Chen-Chen-Meng-ST-14}
X.~Chen, H.~H. Chen, and W.~Meng, ``Cooperative communications for cognitive
  radio networks -- {F}rom theory to applications,'' vol.~16, no.~3, pp.
  1180--1192, 3rd Quart., 2014.

\bibitem{Liu-Others-COMML-13}
D.~Liu \emph{et~al.}, ``Stackelberg game based cooperative user relay assisted
  load balancing in cellular networks,'' \emph{{IEEE} Commun. Lett.}, vol.~17,
  no.~2, pp. 424--427, Feb. 2013.

\bibitem{Zhang-Others-CM-17}
Y.~Zhang \emph{et~al.}, ``A social-aware framework for efficient information
  dissemination in wireless ad hoc networks,'' \emph{{IEEE} Commun. Mag.},
  vol.~55, no.~1, pp. 174--179, Jan. 2017.

\bibitem{L_Lv-Others-TVT-16}
Lv, L. and others, ``Application of
  non-orthogonal multiple access in cooperative spectrum-sharing networks over
  {N}akagami-m fading channels,'' \emph{{IEEE} Trans. Veh. Technol.},
  \mbox{doi}:\url{10.1109/TVT.2016.2627559}.

\bibitem{Z_Zhang-Others-TVT-16}
Z.~Zhang \emph{et~al.}, ``Full-duplex device-to-device aided cooperative
  non-orthogonal multiple access,'' \emph{{IEEE} Trans. Veh. Technol.},
  vol.~66, no.~5, pp. 4467--4471, Aug. 2016.

\end{thebibliography}

\section*{Biographies}
\vspace*{-2\baselineskip}
\begin{IEEEbiographynophoto}{\textsc{Georgios I. Tsiropoulos}}
 (gitsirop@mail.ntua.gr) received the Diploma, M.Sc., and Ph.D. degrees in electrical and computer engineering from National Technical University of Athens, Greece, in 2005, 2009, and 2010, respectively. He served with the Ministry of Transport and Communications of the Hellenic Republic (2006-2009) and the UHC, Greece, (2011-2015) as an ICT consultant. Since 2015, he has been an ICT project manager at the Hellenic Electricity Distribution Network Operator S.A. His current research activities focus on cooperative communications and resource optimization in 5G systems.\vfill
\end{IEEEbiographynophoto} 
\vspace*{-3\baselineskip}

\begin{IEEEbiographynophoto}{\textsc{Animesh Yadav}}
 (animeshy@mun.ca) is a postdoctoral fellow at Memorial University, Canada. Previously, he worked as a postdoctoral fellow and research scientist at UQAM, Canada and CWC at University of Oulu, Finland. During 2003-07, he worked as a software specialist at iGate Inc., India. He received his Ph.D. degree from University of Oulu, Finland. He received the best paper award at IEEE WiMOB-2016. His research interests include enabling technologies for future wireless networks and green communications.\vfill
\end{IEEEbiographynophoto}

\IEEEtriggercmd{}
\begin{IEEEbiographynophoto}{\textsc{Ming Zeng}}
 (mzeng@mun.ca) received his B.E. and master’s degrees from Beijing University of Post and Telecommunications, China. He is now a Ph.D. student in the Faculty of Engineering at Memorial University, Canada. His research interests are in the area of resource allocation in emerging 5G systems.\vfill
\end{IEEEbiographynophoto} 

\begin{IEEEbiographynophoto}{\textsc{Octavia A. Dobre}}
 (odobre@mun.ca) is a Professor and Research Chair at Memorial University, Canada. In 2013 she was a Visiting Professor at Massachusetts Institute of Technology, USA, and University of Brest, France. Previously, she was with New Jersey Institute of Technology, USA, and Polytechnic Institute of Technology, Romania. She was the recipient of a Royal Society scholarship and a Fulbright fellowship. Her research interests include enabling technologies for 5G, cognitive radio systems, blind signal identification and parameter estimation techniques, as well as optical and underwater communications. Dr. Dobre is the EiC of the IEEE Communications Letters. She has served as editor for various prestigious journals, and technical co-chair of different international conferences, such as IEEE ICC and Globecom. She is a member-at-large of the Administrative Committee of the IEEE Instrumentation and Measurement Society, and served as chair and co-chair of different technical committees.\vfill
\end{IEEEbiographynophoto}
\balance

\newpage
\onecolumn


\begin{figure}
\centering
\includegraphics[scale=0.57]{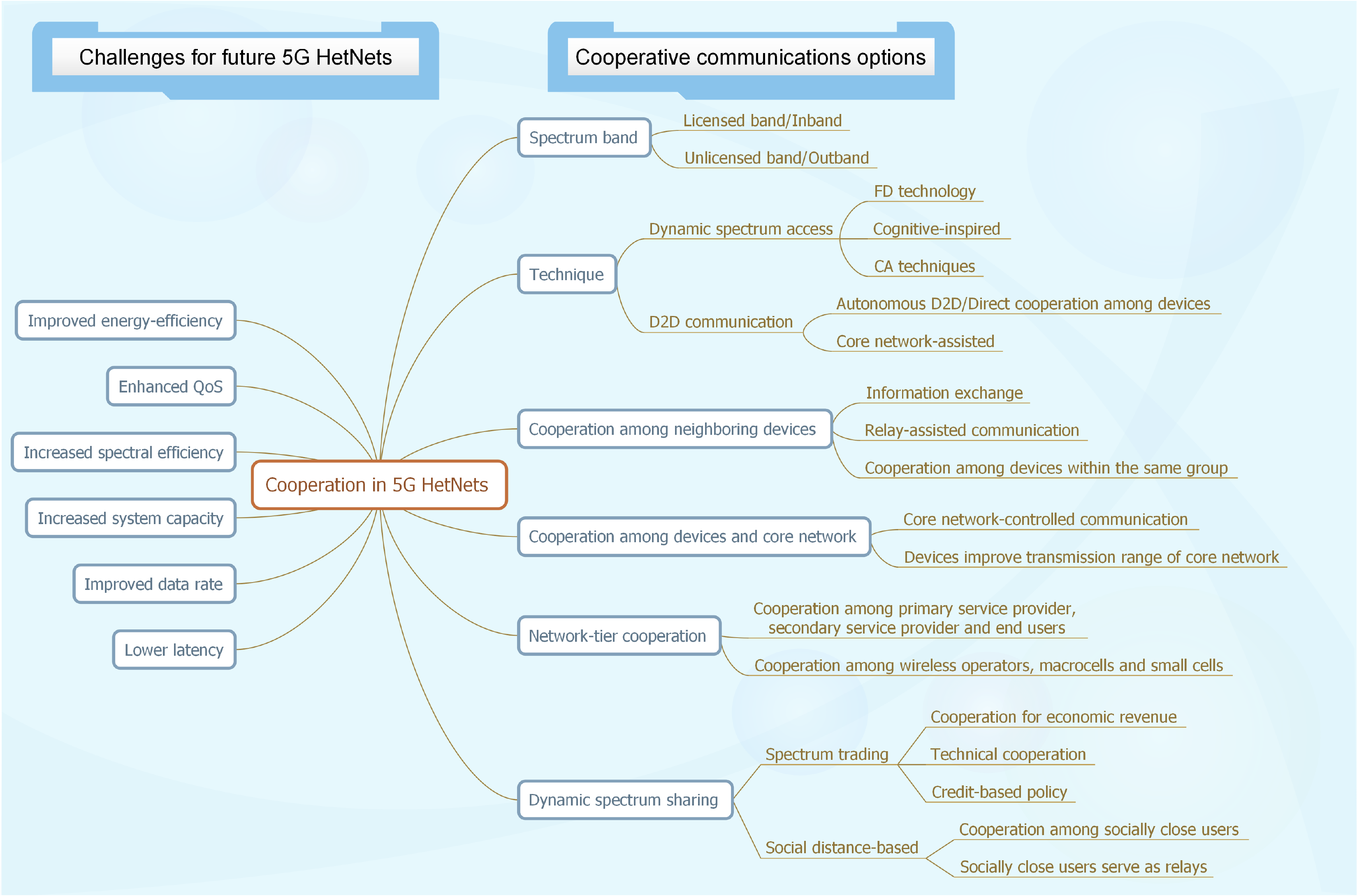}
\caption{Schematic representation of cooperation in 5G HetNet.}\label{fig:Cellular_System}
\end{figure}

\newpage
\begin{figure}
\centering
\includegraphics[scale=0.19]{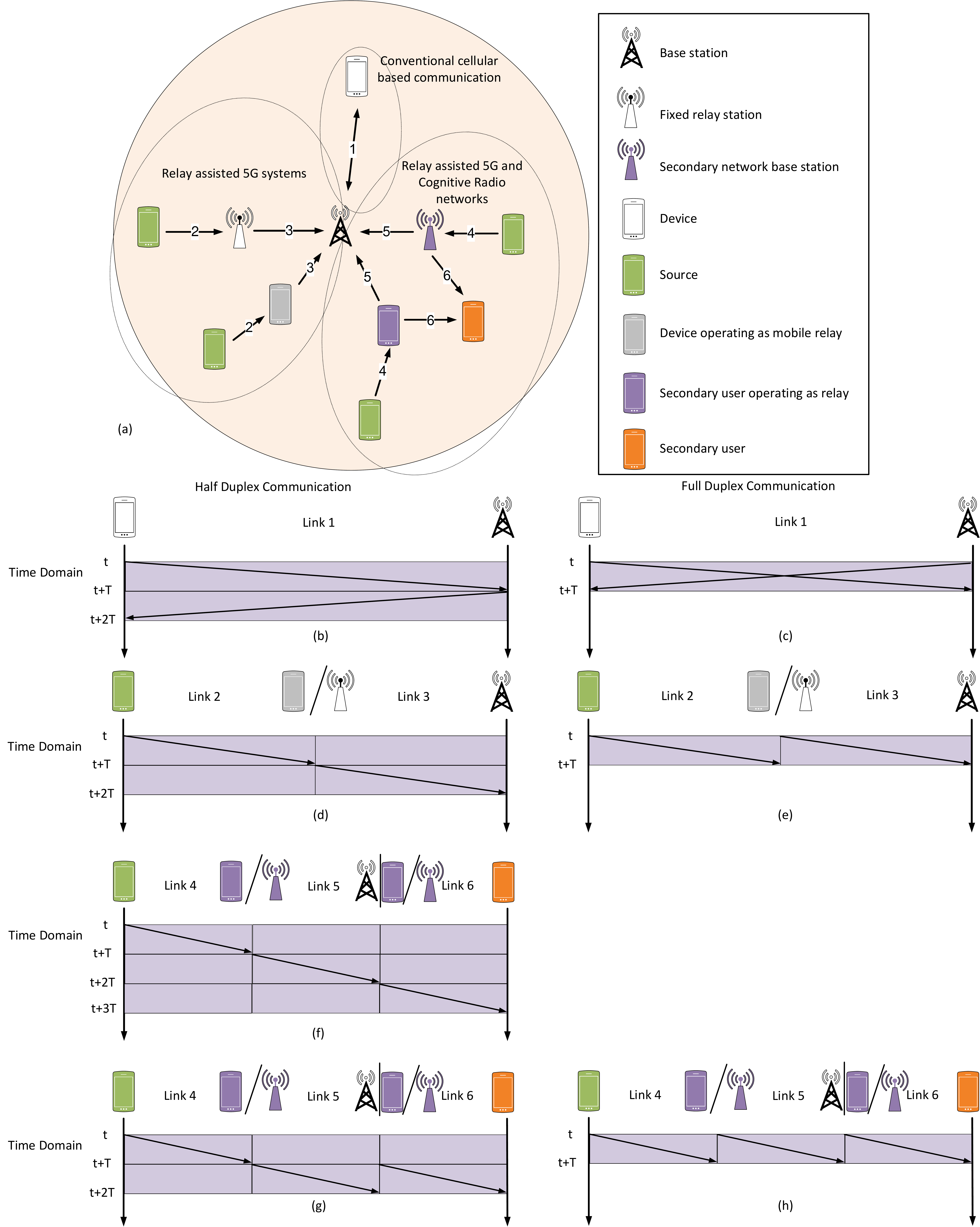}
\caption{Schematic representation of HD vs. cooperative FD spectrum access techniques: (a) several communication paradigms employing FD in 5G HetNet; (b) conventional communication scenario employing HD; (c) conventional communication scenario employing FD; (d) relay-assisted communication employing HD; (e) relay-assisted communication employing FD; (f) relay-assisted communication with 5G and CR networks coexistence employing HD ; (g) relay-assisted communication with 5G and CR networks coexistence employing MIMO and HD; and (h) relay-assisted communication with 5G and CR networks coexistence employing FD.}\label{fig:Cellular_System}

\end{figure}

\newpage

\begin{figure}
\centering
\includegraphics[scale=0.61, trim={0 9cm 0 0}, clip]{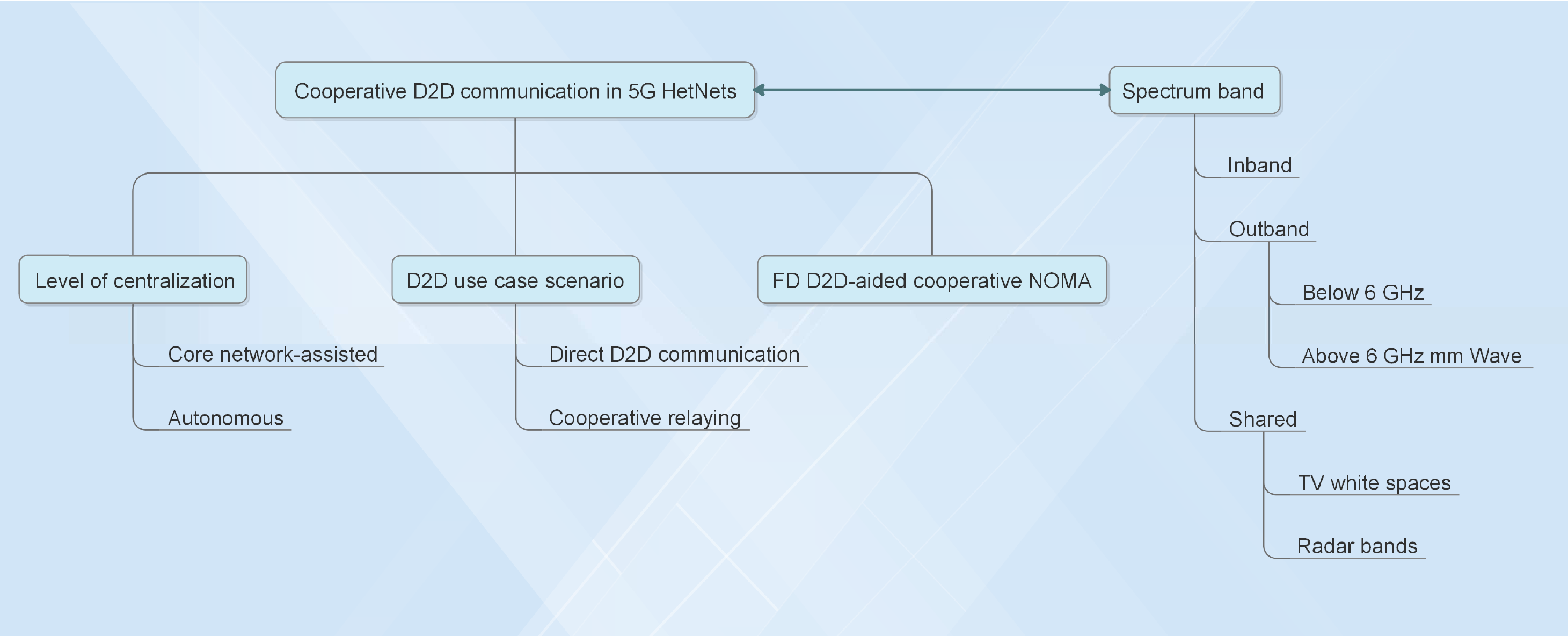}
\caption{A taxonomy of the emerging cooperative D2D communication techniques envisioned for 5G HetNet.}\label{fig:Cellular_System}
\end{figure}

\newpage
\clearpage

\begin{figure}
\centering
\includegraphics[scale=0.19]{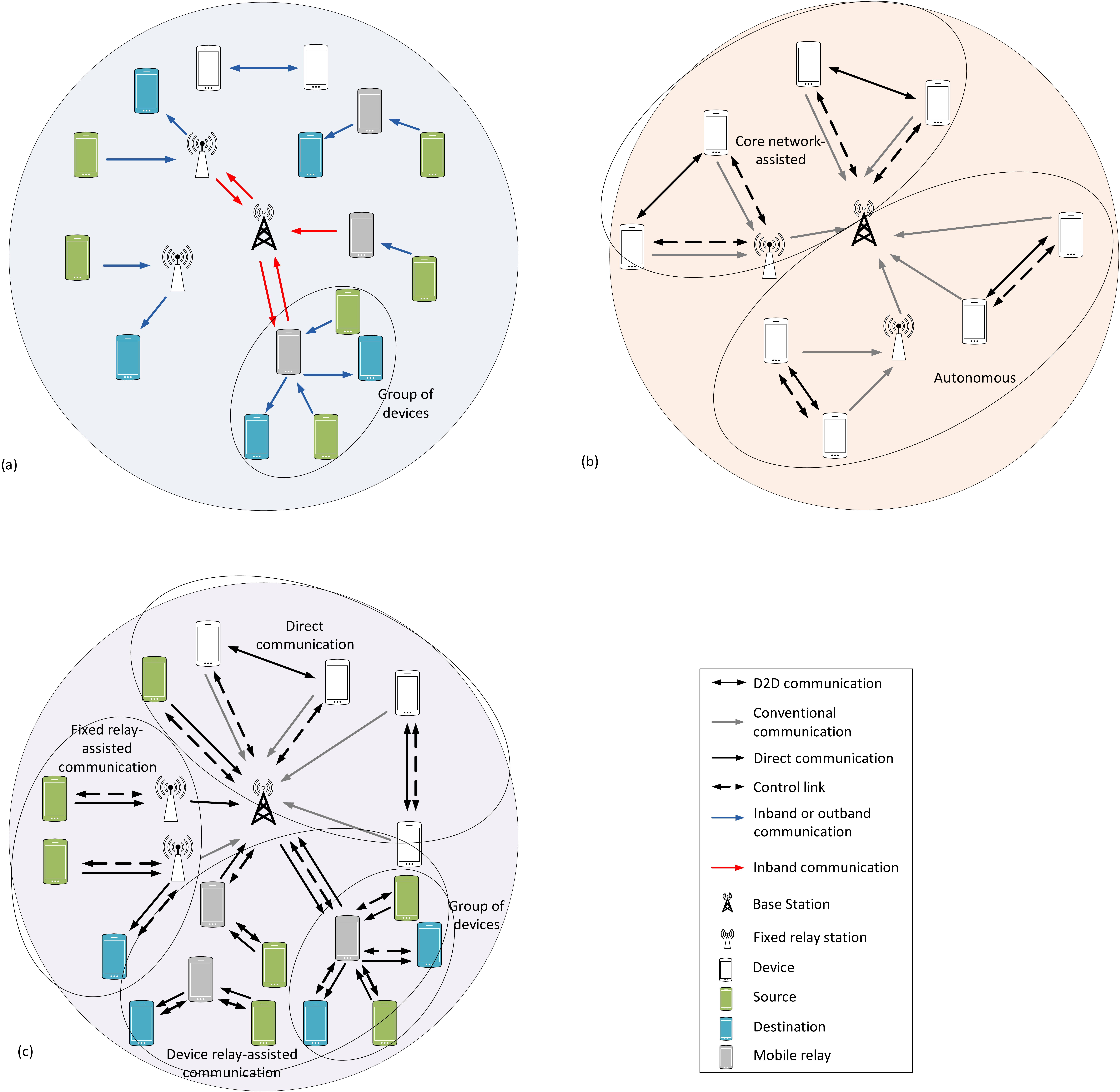}
\caption{Cooperative paradigms on D2D communications in 5G HetNet; (a) inband or outband spectrum utilization; (b) core network-assisted vs. autonomous; and (c) direct vs. relay-assisted communications.}\label{fig:Cellular_System}
\end{figure}

\newpage
\begin{figure}
\centering
\includegraphics[scale=0.19]{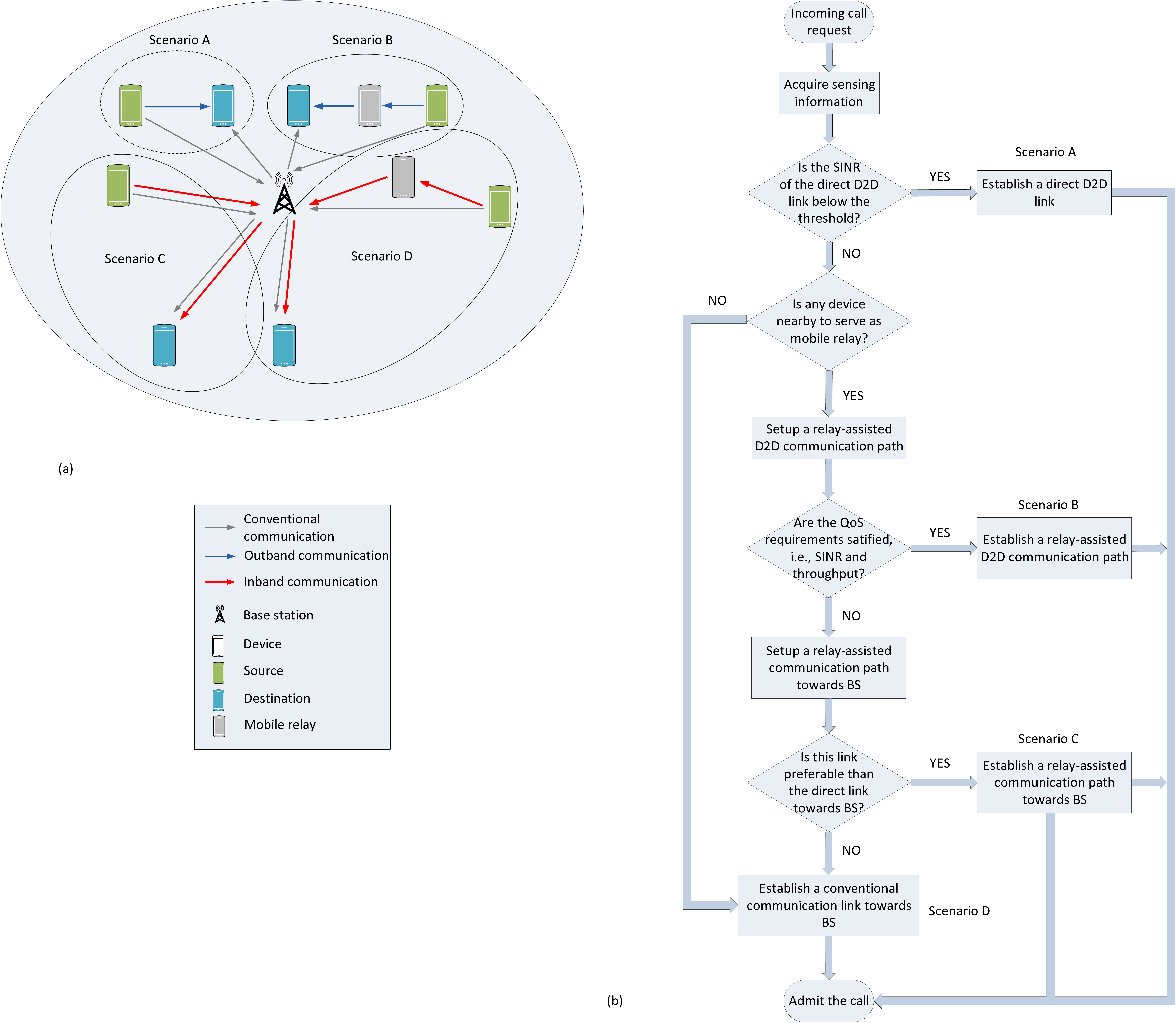}
\caption{The cooperative D2D-assisted model for 5G HetNet: a) four use case scenarios; and b) proposed algorithm for the selection among available communication alternatives.}\label{fig:Cellular_System}
\end{figure}

\newpage

\begin{figure}
\centering
\includegraphics[scale=0.61]{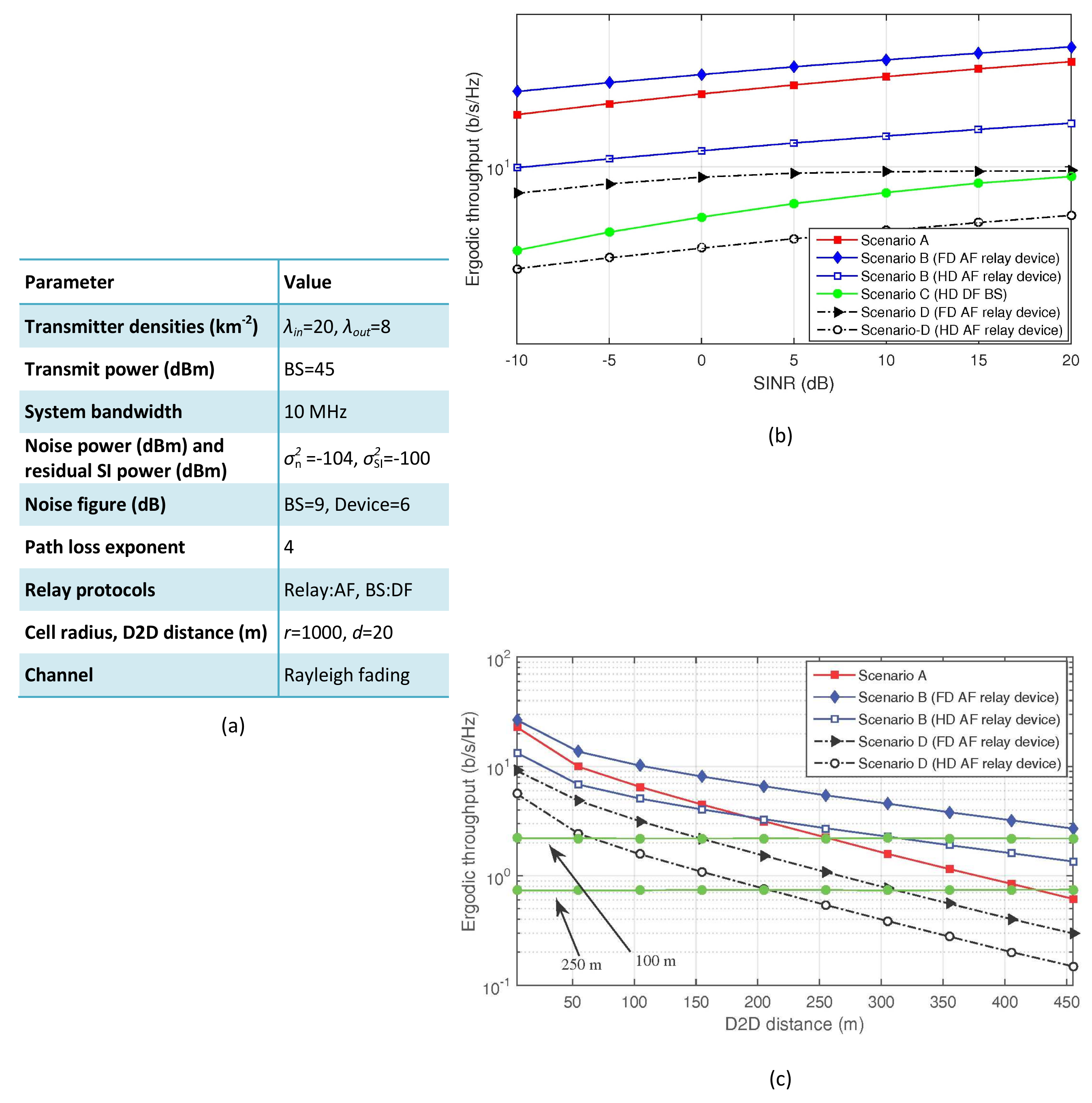}
\caption{Comparison of ergodic throughput performance for different use case scenarios in cooperative 5G HetNet: a) input parameter values; b) ergodic throughput vs. SINR threshold; and c) ergodic throughput vs. D2D distance. Note that $\sigma^{2}_\text{n}$ and $\sigma^{2}_\text{SI}$ represent the noise power and residual SI power, respectively.}\label{fig:Cellular_System}
\end{figure}

\end{document}